\documentclass[prb,superscriptaddress,showpacs,floats]{revtex4}
\usepackage{amsfonts}
\usepackage{amsmath}
\usepackage{amssymb}
\usepackage{graphicx}

\def\t2g{$t_{2g}$}

\input{tcilatex}

\begin{document}

\title{Orbital selective local moment formation in iron: first principle
route to an effective model}
\author{A.~A.~Katanin, A.~I.~Poteryaev, A.~V.~Efremov, A.~O.~Shorikov,
S.~L.~Skornyakov, M.~A.~Korotin, and V.~I.~Anisimov}
\affiliation{Institute of Metal Physics, Russian Academy of Sciences, 620041,
Yekaterinburg GSP-170, Russia}
\pacs{PACS}

\begin{abstract}
We revisit a problem of theoretical description of $\alpha $-iron. By
performing LDA+DMFT calculations in the paramagnetic phase we find that
Coulomb interaction and, in particular Hund exchange, yields the formation
of local moments in $e_{g}$~electron band, which can be traced from
imaginary time dependence of the spin-spin correlation function. This
behavior is accompanied by non-Fermi-liquid behavior of $e_{g}$ electrons
and suggests using local moment variables in the effective model of iron. By
investigating orbital-selective contributions to the Curie-Weiss law for
Hund exchange $I=0.9$~eV we obtain an effective value of local moment of $%
e_{g}$ electrons $2p=1.04\mu _{B}$. The effective bosonic model, which
allows to describe magnetic properties of iron near the magnetic phase
transition, is proposed.
\end{abstract}

\date{date}
\maketitle

\section{Introduction}

\label{sec:intro}

The magnetism and its influence to properties of materials attracts a lot of
interest since ancient ages, first records can be traced back to Greek
philosopher Thales of Miletus and Indian surgeon Sushruta about 600~BC. In
particular, the problem of origin of ferromagnetism of iron attracts a lot
of attention, despite long time of its investigations.

The $d$-electrons in iron (as well as in many other transition metals) show
both, localized, and itinerant behavior. According to the Rhodes and
Wolfarth criterion, iron is classified as a local moment system, since the
ratio of the magnetic moment $2p_{\text{CW}}\mu _{B}$ corresponding to the
effective spin $p_{\text{CW}}$ extracted from Curie-Weiss law for
susceptibility $\chi =g^{2}\mu _{B}^{2}p_{\text{CW}}(p_{\text{CW}%
}+1)/(3(T-T_{C}))$ ($g\approx 2$ is the $g$-factor, $\mu _{B}$ is the Bohr
magneton, $T_{C}$ being the Curie temperature) to the magnetic moment per
atom in the ferromagnetic phase, $2p_{\text{CW}}\mu _{B}/\mu _{\text{exp}%
}=1.05$ is close to unity (see, e.g. Ref.~\onlinecite{Moriya}). At the same
time, the experimental magnetic moment of iron $\mu _{\exp }=2.2\mu _{B}$ is
not an integer number, which indicates presence of some fraction of
itinerant electrons.

Itinerant theory of magnetism of transition metals was pioneered by Stoner,
and then became a basis of spin-fluctuation theory by Moriya~\cite{Moriya}
which was successful to describe weak and nearly ferro- and
antiferromagnetic materials. By considering fluctuation corrections to mean
field, Moriya theory was able to reproduce nearly Curie-Weiss behavior of
magnetic susceptibility and to obtain correct values of transition
temperatures of weak or nearly magnetic systems. At the same time, this
theory meets serious difficulties when applied to materials with large
magnetic moment, such as some transition metals. These materials are
expected to be better described in terms of the local moment picture. In
practice, to describe $d$-electrons in transition metals in the
semi-phenomenological way, the localized-moment (Heisenberg) model is often
used. Band structure calculations of magnetic exchange interaction in iron
show however its non-Heisenberg character at intermediate and large momenta~%
\cite{Licht}. Using microscopic consideration Mott~\cite{Mott} proposed a
two-band model for transition metals with narrow band of $d$-electrons and
wide band of $s$-electrons. The polar $s$-$d$ model, which treats $d$%
-electrons as localized and $s$-electrons as itinerant was proposed by
Shubin and Vonsovskii~\cite{Shubin}.

First attempts to unify the localized and itinerant pictures of magnetism
were performed in Refs.~\onlinecite{Hubbard,Moriya} for the single and
degenerate band models, respectively. To unify localized and itinerant
approaches to magnetism and find an origin of the formation of local
moments, it seems however important to consider the orbital-resolved
contributions to one- and two-particle properties. In particular, it was
suggested by Goodenough~\cite{Goodenough} that the the electrons with $e_{g}$
and $t_{2g}$ symmetry may behave very differently in iron: while the former
show localized, the latter may show itinerant behavior. The
\textquotedblleft 95\% localized model\textquotedblright\ of iron was
proposed by Stearns~\cite{Stearns} according to which 95\% of $d$-electrons
are localized, while 5\% are itinerant. This idea found its implementation
in the \textquotedblleft two-band model\textquotedblright\ \cite{Mota},
which was considered within the mean-field approach. Later on it was
suggested \cite{KatsVH,KatsVH1} that the states at the van Hove
singularities may induce localization of some $d$-electron states. However,
no microscopic evidences for such localization were obtained so far.

The important source\ of the local moment formation are strong electronic
correlations. In particular, the ferromagnetic state of the one-band
strongly-correlated Hubbard model, which was shown to be stable for
sufficiently large on-site Coulomb repulsion\cite{Uhrig,DMFTFerro}, has
linear dependence of the inverse susceptibility above transition
temperature~within the dynamical mean-field theory (DMFT)\cite{DMFTFerro}.
The role of interband Coulomb interaction and Hund exchange in
non-degenerate Hubbard model to reduce the critical intraband Coulomb
interaction strength was emphasized in Refs.~%
\onlinecite{DMFTFerroMultiorb,Rio}.

To get insight in the applicability of the abovementioned proposals to
mechanism of local moment formation in iron, the combination of
first-principle\cite{Manning,Callaway,Abate} and model calculations seems
necessary. The recently performed LDA+DMFT calculations\cite{KatsPRL}
allowed to describe quantitatively correct the magnetization and
susceptibility of iron as a function of the reduced temperature $T/T_{C};$
in particular they led to almost linear temperature dependence of the
inverse static spin susceptibility above the magnetic transition
temperature, which is similar to the results of model calculations and can
be considered as possible evidence for existence of local moments. The
estimated magnetic transition temperature appears however twice large than
the experimental value $T_{C}=1043K$. The one-particle properties below the
transition temperature were addressed in Refs.~%
\onlinecite{MagnDMFT,KatsLicht}. To get insight into the mechanism of the
formation of local moments and linear dependence of susceptibilities above
the Curie temperature it seems however important to study one- and
two-particle properties in the \textit{symmetric} phase. 

To this end we reconsider in the present paper ab initio LDA+DMFT
calculations, paying special attention to orbital-resolved contributions to
one- and two-particle properties. Contrary to previously accepted view that
Hund exchange only helps to form ferromagnetic state, we argue that in fact
it serves as a main source of formation of local moments in iron, together
with the almost absent hybridization between $t_{2g}$ and $e_{g}$ bands.
These two factors yield formation of local moments for the $e_{g}$ states,
while $t_{2g}$ states remain more itinerant.

\section{The $d$-electron model and orbital-selective magnetic moments}

To discuss the behavior of $d$-electrons in iron let us start from standard
multi-band Hubbard Hamiltonian 
\begin{align}
\hat{H}^{d}& =\hat{H}_{\text{kin}}+\hat{H}_{\text{int}}^{d}
\label{eq:Hubbard} \\
+& \sum_{\mathbf{k}}\sum\limits_{mm^{\prime }\sigma }t_{mm^{\prime }}(%
\mathbf{k})\hat{c}_{\mathbf{k}m\sigma }^{\dagger }\hat{c}_{\mathbf{k}%
m^{\prime }\sigma }  \notag \\
+& \frac{1}{2}\sum_{i}\sum\limits_{\{m\}\sigma \sigma ^{\prime }}\langle
m,m^{\prime }|V_{ee}|m^{\prime \prime },m^{\prime \prime \prime }\rangle 
\hat{c}_{im\sigma }^{\dagger }\hat{c}_{im^{\prime }\sigma ^{\prime
}}^{\dagger }\hat{c}_{im^{\prime \prime }\sigma }\hat{c}_{im^{\prime \prime
\prime }\sigma ^{\prime }},  \notag
\end{align}%
where the first term represents a kinetic contribution to Hamiltonian and
the second one is an interaction part. $c_{\mathbf{k}m\sigma }^{\dagger }(c_{%
\mathbf{k}m\sigma })$ are creation (annihilation) operators for electron
with respective quantum indices $|\mathbf{k}m\sigma \rangle $ and $%
c_{im\sigma }^{\dagger }$ is a Fourier image in real space. $t_{mm^{\prime
}}(\mathbf{k})$ is a dispersion and $\langle m,m^{\prime }|V_{ee}|m^{\prime
\prime },m^{\prime \prime \prime }\rangle $ is a Coulomb interaction matrix.
For a sake of simplicity we assume that orbital index $m$ runs over the
correlated $d$-orbitals only.

Keeping a density-density and spin-flip terms in the interacting part of
above Hamiltonian (Eq.~\ref{eq:Hubbard}) and assuming a simple
parametrization of the interaction matrix with the intraorbital Coulomb
interaction, $U=\langle m,m|V_{ee}|m,m\rangle $, the interorbital Coulomb
interaction, $U^{\prime }=\langle m,m^{\prime }|V_{ee}|m,m^{\prime }\rangle $
and Hund's exchange, $I=\langle m,m^{\prime }|V_{ee}|m^{\prime },m\rangle $
one can rewrite the interaction as %
\begin{align}
H_{\text{int}}^{d}& =U\sum\limits_{im}\hat{n}_{im\uparrow }\hat{n}%
_{im\downarrow }+(U^{\prime }-\frac{I}{2})\sum\limits_{i,m<m^{\prime
},\sigma \sigma ^{\prime }}\hat{n}_{im\sigma }\hat{n}_{im^{\prime }\sigma
^{\prime }}  \label{eq:hintmodel} \\
& -2I\sum\limits_{i,m<m^{\prime }}\hat{\mathbf{s}}_{im}\hat{\mathbf{s}}%
_{im^{\prime }},  \notag
\end{align}%
where $m$ runs over all $d$-orbital indices and 
\begin{align*}
\hat{n}_{im\sigma }& =\hat{c}_{im\sigma }^{\dagger }\hat{c}_{im\sigma }, \\
\hat{\mathbf{s}}_{im}& =\frac{1}{2}\sum\limits_{\sigma \sigma ^{\prime }}%
\hat{c}_{im\sigma }^{\dagger }\boldsymbol{\sigma}_{\sigma \sigma ^{\prime }}%
\hat{c}_{im\sigma ^{\prime }}.
\end{align*}%
$\boldsymbol{\sigma}$ are the Pauli matrices.

Generically, the Coulomb interaction yields loss of coherence of
corresponding electronic states. It will be shown in Sec.~\ref%
{sec:calculations}, that electrons in weakly hybridized $t_{2g}$ and $e_{g}$
orbitals behave very differently with respect to the Coulomb interaction.
While the behavior of $t_{2g}$ electrons remains Fermi liquid like, $e_{g}$
electrons form a non-Fermi liquid states that implies formation of local
moments. Magnetic properties of the resulting system can be then understood
in terms of an effective model, containing spins of local and itinerant
electron subshells.

Splitting in Eq.~(\ref{eq:hintmodel}) contributions of $e_{g}$ and $t_{2g}$
electrons and neglecting hybridization between them (which will be shown to
be small in Sec.~\ref{sec:lda}), we can rewrite the Hamiltonian (\ref%
{eq:Hubbard}) as 
\begin{align}
\hat{H}^{d}& =\hat{H}^{t_{2g}}+\hat{H}^{e_{g}}-2I\sum\limits_{i,m\in t_{2g}}%
\hat{\mathbf{S}}_{i}\hat{\mathbf{s}}_{im}  \label{Hdd} \\
& +(U^{\prime }-\frac{I}{2})\sum\limits_{i,\sigma ,m\in t_{2g}}\hat{N}_{i}%
\hat{n}_{im\sigma },  \notag
\end{align}%
where $\hat{H}^{t_{2g}}$ and $\hat{H}^{e_{g}}$ are the parts of Hamiltonian~(%
\ref{eq:Hubbard}) acting on the $t_{2g}$ and $e_{g}$ orbitals, respectively, 
$\hat{\mathbf{S}}_{i}=\sum\nolimits_{m\in e_{g}}\hat{\mathbf{s}}_{im},~\hat{N%
}_{i}=\sum\nolimits_{m\in e_{g}}\hat{n}_{im}$. Note that operators $\hat{%
\mathbf{S}}_{i}$ do not generically describe fully local moments, but will
be shown to have properties close to those of local moments due to Hund
exchange interaction. Below after considering the results of band structure
calculations, we discuss the effect of interaction in Eq.~(\ref{Hdd}) within
DMFT and its implications for the effective model.

\section{First principle calculations for iron}

\label{sec:calculations}

\subsection{Band structure results}

\label{sec:lda}

\begin{figure}[tbp]
\centering              
\includegraphics[clip=true,width=0.33\textwidth,
angle=270]{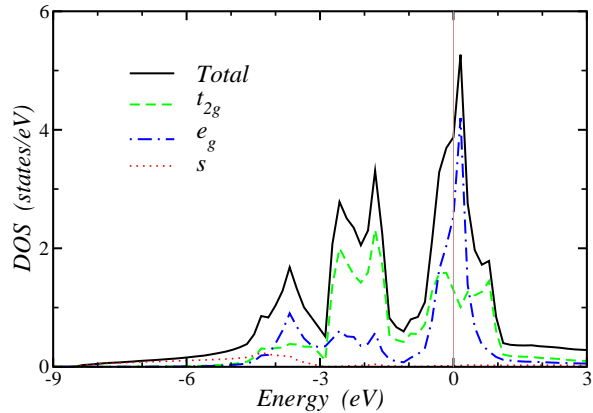}
\caption{(Color online) Iron density of states obtained within LDA
approximation. Total DOS is shown by solid (black) line. Partial \t2g, $%
e_{g} $ and $s$ DOSes are shown by (green) dashed, (blue) dot-dashed and
(red) dot lines, respectively.}
\label{fig:lda_dos}
\end{figure}

\begin{figure*}[!hbt]
\centering               
\includegraphics[clip=true, width=0.33\textwidth,
angle=270]{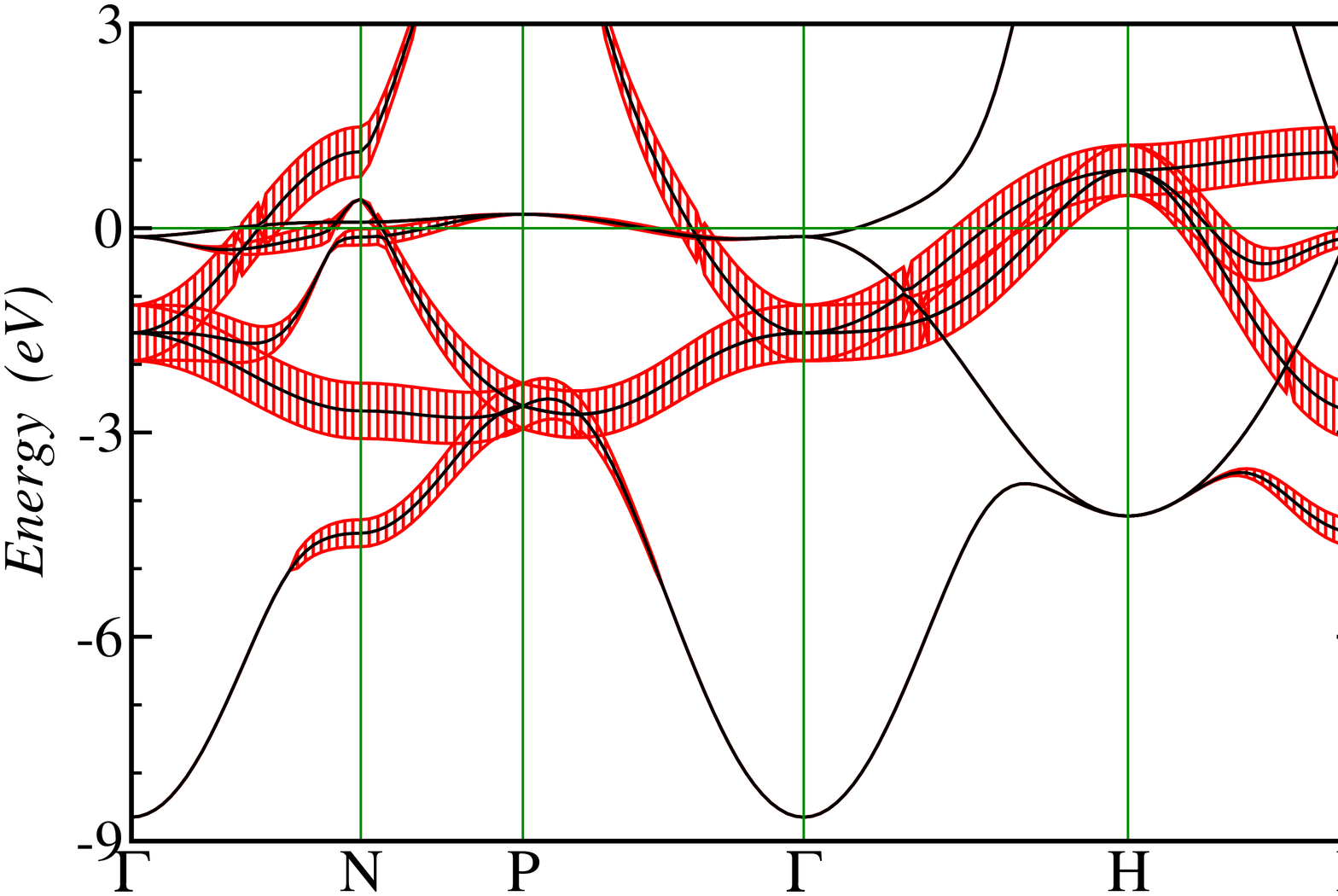} 
\includegraphics[clip=true, width=0.33\textwidth,
angle=270]{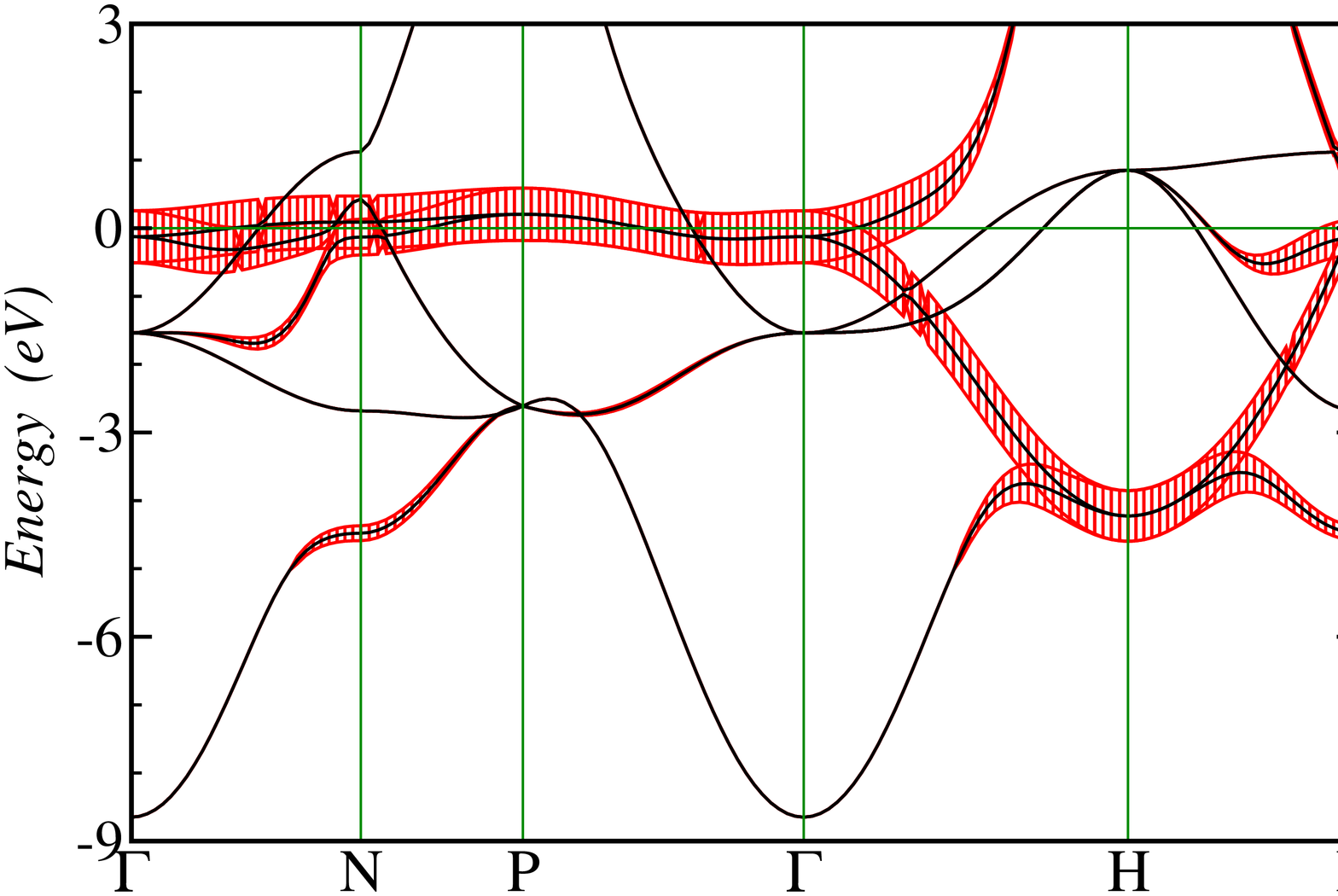}
\caption{(Color online) The band structure of Fe along high symmetry lines
in the Brillouin zone obtained within LDA approximation. The contribution of 
$t_{2g}$ (left panel) and $e_{g}$ (right panel) states is shown with fat
(red) lines.}
\label{fig:fat_band}
\end{figure*}

Iron crystallizes in body centered cubic structure below 1183~K and has the
lattice parameter $a=2.8664$~\AA\ at room temperature~\cite{str}. Band
structure calculations have been carried out in LDA approximation~\cite{LDA}
within TB-LMTO-ASA framework~\cite{Andersen84}. The von Barth-Hedin local
exchange correlation potential was used~\cite{vonbarth}. Primitive
reciprocal translation vectors were discretized into 12 points along each
direction which leads to 72~$\mathbf{k}$-points in irreducible part of the
Brillouin zone.

Total and partial densities of states are presented in Fig.~\ref{fig:lda_dos}%
. The contribution of a wide $s$-band is shown by (red) dots and spreads
from -8.5~eV to energies well above the Fermi level (at zero energy); $t_{2g}
$ and $e_{g}$~states ((green) dashed and (blue) dot-dashed) span energy
region from -5~eV to 1~eV approximately. In spite of almost equal bandwidths
of $t_{2g}$ and $e_{g}$~states they are qualitatively different. Former
states are distributed more uniformly over the energy range while the later
one have a large peak located at the Fermi energy.

The contributions of $t_{2g}$ and $e_{g}$ orbitals to the band structure of
iron are presented in Fig.~\ref{fig:fat_band} (left and right panels,
respectively). The states contributing to the van Hove singularity near the
Fermi energy are of mostly $e_{g}$ symmetry. As it was argued in Refs.~%
\onlinecite{KatsVH1} and~\onlinecite{VHS}, despite the three-dimensional
character of the band structure, the lines of van Hove singularities, which
due to symmetry reasons can easily occur along the $\Gamma -N$ direction,
produce a peak in the density of states. In fact, this singularity is
actually a part of the flat band going along $\Gamma -N-P-\Gamma $
directions. On the other hand, $t_{2g}$ bands do not have a flatness close
to the Fermi level. These peculiarities of the band structure and absence of
direct hybridization between $t_{2g}$ and $e_{g}$ states suggest that the $%
t_{2g}$ and $e_{g}$~electrons may behave very differently when turning on
on-site Coulomb interaction.

\subsection{DMFT calculations}

\label{sec:dmft}

\begin{figure}[tbh]
\centering              
\includegraphics[clip=true,
width=0.33\textwidth,angle=270]{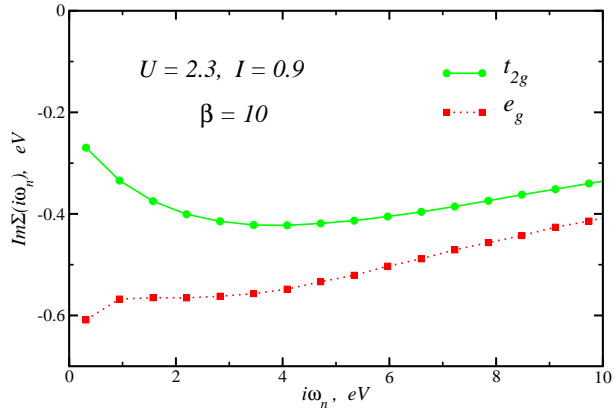}
\caption{(Color online) The imaginary part of self-energy for $t_{2g}$
(green) solid line and $e_{g}$ states (red) dotted line plotted on the
Matsubara (imaginary) energy grid.}
\label{fig:sigma_im}
\end{figure}

In order to take into account correlation effects in 3$d$ shell of $\alpha $%
-iron the LDA+DMFT method was applied (for detailed description of the
computation scheme see Ref.~\onlinecite{Anisimov05}). We use the Hamiltonian
of Hubbard type as in Eq.~(\ref{eq:Hubbard}) with the kinetic term
containing all $s-p-d$ states and the interaction part with density-density
contributions for $d$-electrons only 
\begin{align}
\hat{H}_{\text{int}}& =\frac{1}{2}\sum\limits_{imm^{\prime }\sigma
}\Bigl\{U_{mm^{\prime }}\hat{n}_{im\sigma }\hat{n}_{im^{\prime }\bar{\sigma}}
\label{eq:h_int_denden} \\
& +(U_{mm^{\prime }}-J_{mm^{\prime }})\hat{n}_{im\sigma }\hat{n}_{im^{\prime
}\sigma }\Bigr\},  \notag
\end{align}%
where $U_{mm^{\prime }}\equiv \langle m,m^{\prime }|V_{ee}|m,m^{\prime
}\rangle $ and $J_{mm^{\prime }}\equiv \langle m,m^{\prime
}|V_{ee}|m^{\prime },m\rangle $. Regarding interaction between $d$%
-electrons, the model (\ref{eq:h_int_denden}) serves as a simplified version
of the model~(\ref{eq:Hubbard}), since it does not contain transverse
components of the Hund exchange and pair-hopping term.

The Coulomb interaction parameter value $U$=2.3~eV and the Hund's parameter $%
I$=0.9~eV used in our work are the same as in earlier LDA+DMFT calculations
by Lichtenstein \textit{et al}~\cite{KatsPRL}. To treat a problem of
formation of local moments we consider paramagnetic phase. The effective
impurity model for DMFT was solved by QMC method with the Hirsh-Fye
algorithm~\cite{HF86}. Calculations were performed for the value of inverse
temperature $\beta $=10~eV$^{-1}$ which is close to the $\alpha \rightarrow
\gamma $ transition temperature. Inverse temperature interval $0<\tau <\beta 
$ was divided in 100 slices. 4 million QMC sweeps were used in
self-consistency loop within LDA+DMFT scheme and up to 12 million of QMC
sweeps were used to calculate spectral functions.

In Fig.~\ref{fig:sigma_im} the imaginary part of self-energies are shown for
the imaginary frequency axis. One can clearly see that the behavior of $\Im
\Sigma (i\omega _{n})$ at low energies is qualitatively different for
different orbitals. While $\Im \Sigma (i\omega _{n})$ for $t_{2g}$ states
has a Fermi-liquid-like behavior with the quasiparticle weight $Z$=0.86,
zero energy outset $\Re \Sigma (0)\simeq 1.1$~eV, and damping $\Im \Sigma
(0)=-0.22$~eV, the $\Im \Sigma (i\omega _{n})$ for $e_{g}$ orbitals has a
divergent-like shape indicating a loss of coherence regime. As it will be
shown in the Section~\ref{sec:dmft_chi}, the latter states form local
magnetic moments. This fact affords a ground for separation of the iron $d$%
-states onto two subsystems: more localized $e_{g}$-states and itinerant $%
t_{2g}$-states. Contrary to the picture proposed in Ref.~\onlinecite{KatsVH}%
, we find not only localization of electrons, contributing to the van Hove
singularity states, but most part of $e_{g}$ electrons is expected to form
local moments. The features observed for $e_{g}$ states are similar to those
observed near Mott metal insulator transition~\cite{Bulla} (see also the
results on the real axis below in Fig.~\ref{fig:sigma_re}), although in our
case non-Fermi liquid behavior touches only part of the states and the
metal-insulator transition does not happen. We have verified that the
obtained results depend very weakly on $U$ in the range $U=2\div 6$~eV,
while switching off (or reducing) $I$ immediately suppresses
non-Fermi-liquid contributions. Therefore, Hund exchange serves as a major
source of local moment formation of $e_{g}$ states.

Fig.~\ref{fig:sigma_re} shows the resulting behavior of real and imaginary
parts of the self-energy of different orbitals on the real axis. To make an
analytic continuation of the complex function we used Pade approximation
method~\cite{pade} with energy mesh containing both, low- and high energy
frequencies. To satisfy high-frequency behaviour the equality of the first
three moments of function calculated on the real and imaginary axis was
fullfilled. Altogether this procedure garantees an accurate description of
the function close to Fermi level and at high-energy. In compliance with the
observations from imaginary axis, $\Re \Sigma (\omega )$ has slightly
negative slope for $t_{2g}$ states, accompanied by the maximum of $\Im
\Sigma (\omega )$ at the Fermi level, while for $e_{g}$ states $\Re \Sigma
(\omega )$ has positive slope and $\Im \Sigma (\omega )$ is minimal at the
Fermi level. The characteristic energy scale for the observed non-Fermi
liquid behavior is of the order of 1~eV, i.e. the Hund exchange parameter,
which is too small to produce Hubbard subbands, see straight lines in Fig.~%
\ref{fig:sigma_re}. Seemingly, the observed features represent stronger
breakdown of the Fermi liquid behavior, than $\Im \Sigma \varpropto
T^{1+\alpha }$ obtained earlier in the three-band Hubbard model~\cite{werner}%
. 

\begin{figure}[tbh]
\centering              
\includegraphics[clip=true, width=0.33\textwidth,
angle=270]{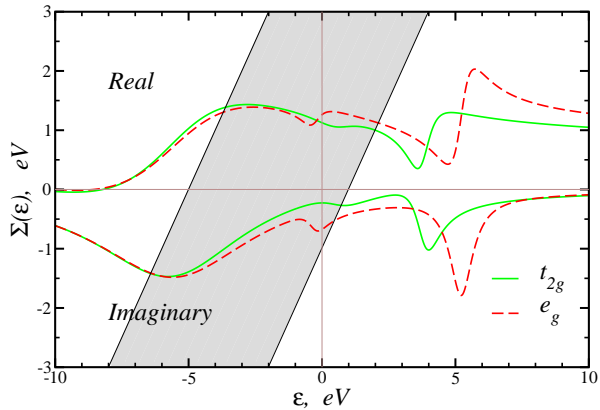}
\caption{(Color online) The self-energy for $t_{2g}$ (green) solid line and $%
e_{g}$ states (red) dashed line plotted on the real energy axis. Straight
lines $\Sigma=\protect\omega+E_{\text{min,max}}$, which bound shaded area,
correspond to the bottom ($E_{\text{min}}$) and top ($E_{\text{max}}$) of
the band.}
\label{fig:sigma_re}
\end{figure}

\begin{figure*}[tbh]
\centering              
\includegraphics[clip=true, width=0.33\textwidth,
angle=270]{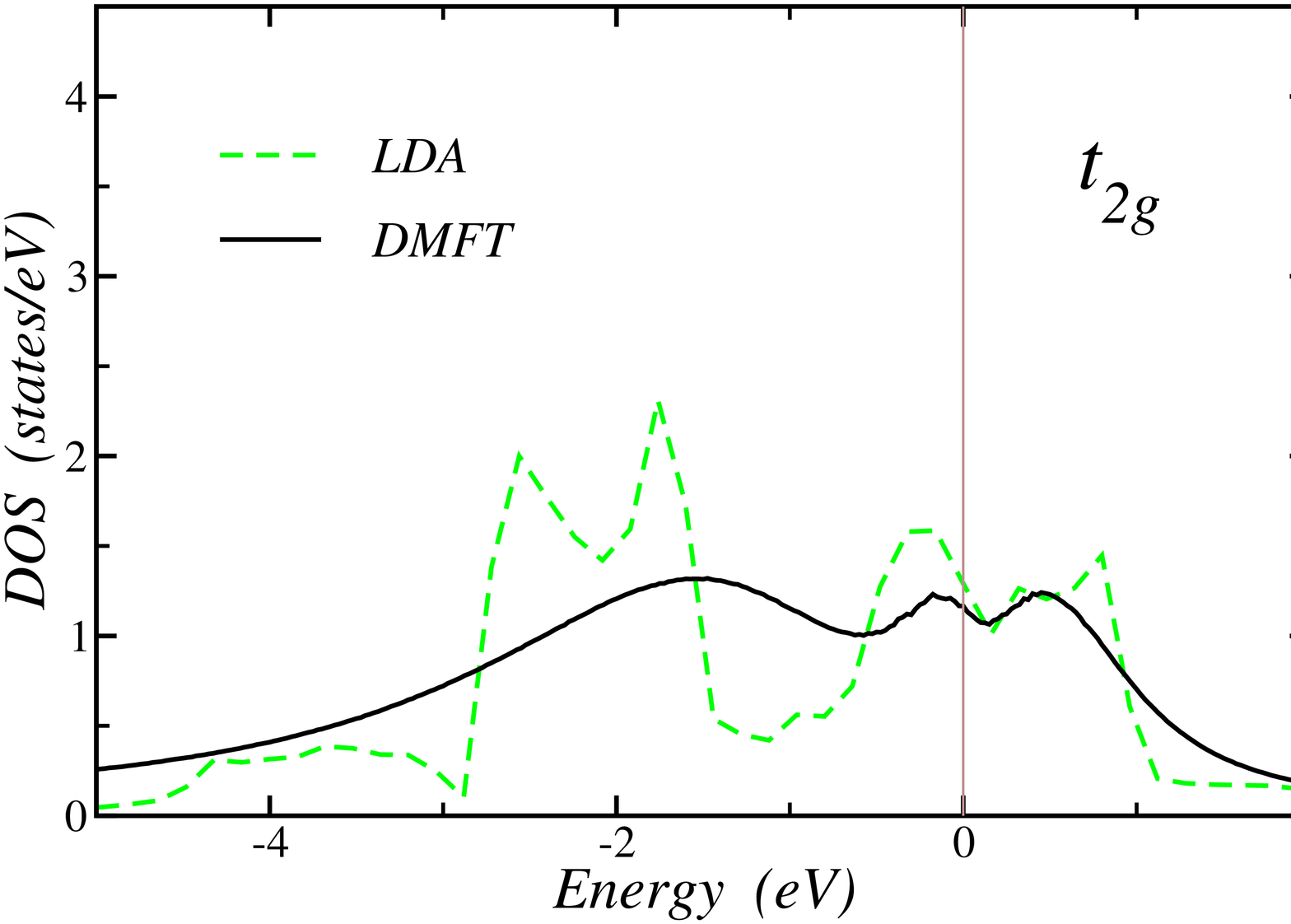} 
\includegraphics[clip=true,
width=0.33\textwidth, angle=270]{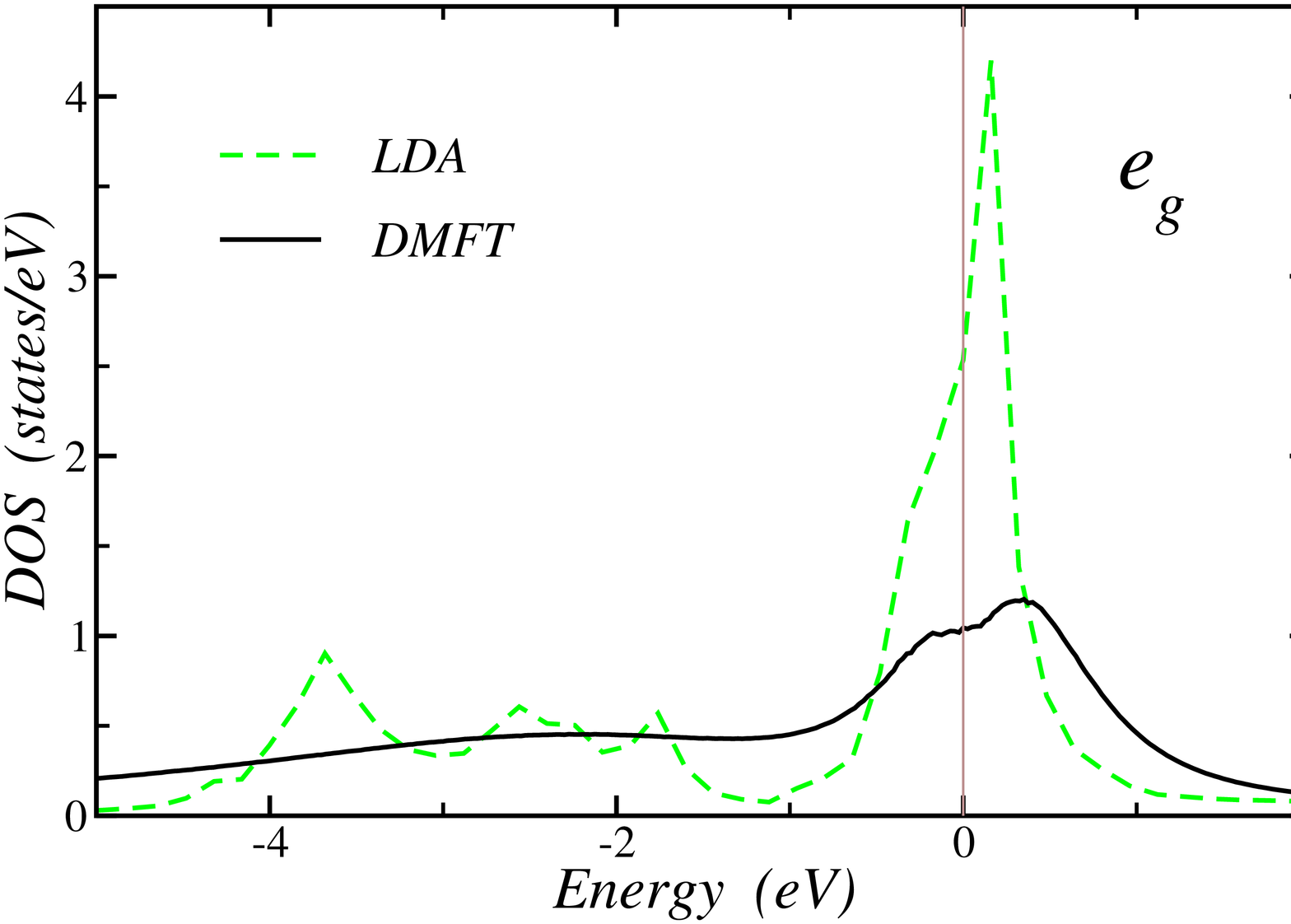}
\caption{(Color online) The iron $t_{2g}$ (left panel) and $e_{g}$ (right
panel) partial density of states obtained within LDA+DMFT method (black
solid lines) compared with LDA DOS (green dashed lines).}
\label{fig:dos_dmft}
\end{figure*}

Partial densities of states obtained in paramagnetic LDA+DMFT calculation
for $t_{2g}$ and $e_{g}$ electrons are presented in Fig.~\ref{fig:dos_dmft}.
The LDA+DMFT densities of states are slightly narrower than the LDA
counterparts implying weak correlation effects. One can observe that peak of 
$e_{g}$ density of states observed in LDA approach is suppressed in LDA+DMFT
calculation and split into two peaks at $-0.3$ and 0.5~eV due to
non-Fermi-liquid behavior of these states. As discussed above, this
splitting should be distinguished from the Hubbard subbands formation near
Mott metal-insulator transition, e.g. due to much smaller energy scale,
which is of the order of Hund exchange interaction. The shape of $t_{2g}$
density of states in LDA+DMFT approach resembles the LDA result with
smearing of the peaky structures of density of states by correlations.

\subsection{DMFT spin susceptibility}

\label{sec:dmft_chi}

\begin{figure}[tbh]
\centering              
\includegraphics[clip=true,
width=0.45\textwidth]{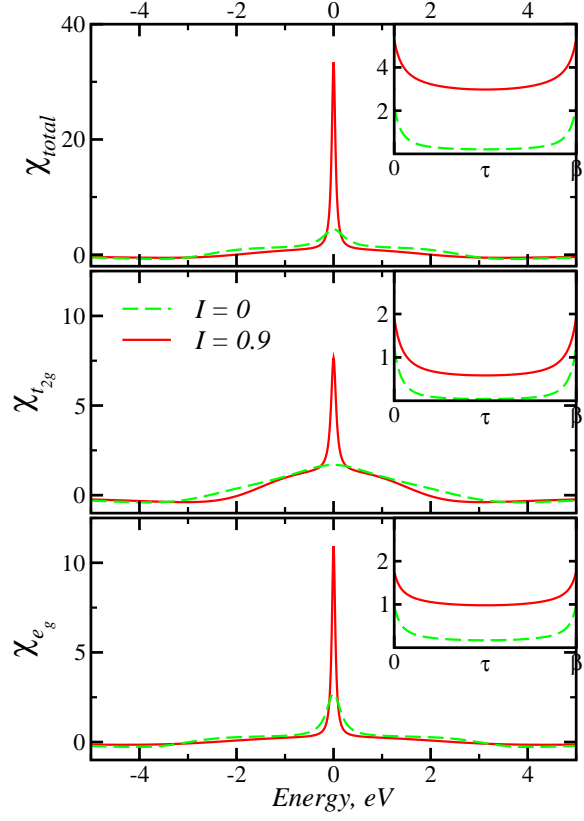}
\caption{(Color online) Impurity spin susceptibility for the value of
Coulomb interaction, $U$ = 2.3 eV, inverse temperature, $\protect\beta$=10 eV%
$^{-1}$ and Hund coupling, $I$=0 (green dashed) and 0.9 eV (red solid)
plotted on the real axis. Total impurity spin susceptibility and \t2g and $%
e_{g}$ contributions are shown from top to bottom. The insets show the
corresponding imaginary time data.}
\label{fig:chi_imp}
\end{figure}

To discuss the effect of the non-quasiparticle states of $e_{g}$ electrons
on magnetic properties we consider imaginary time dependence of the impurity
spin susceptibilities 
\begin{eqnarray*}
\chi _{e_{g}}(\tau ) &=&\sum\limits_{mm^{\prime }\in e_{g}}\langle T[\hat{s}%
_{im}^{z}(\tau )\hat{s}_{im^{\prime }}^{z}(0)]\rangle =\langle T[\hat{S}%
_{i}^{z}(\tau )\hat{S}_{i}^{z}(0)]\rangle  \\
\chi _{t_{2g}}(\tau ) &=&\sum\limits_{mm^{\prime }\in t_{2_{g}}}\langle T[%
\hat{s}_{im}^{z}(\tau )\hat{s}_{im^{\prime }}^{z}(0)]\rangle 
\end{eqnarray*}%
obtained within DMFT. The results for the time dependence of $\chi
_{e_{g}}(\tau )$, $\chi _{t_{2g}}(\tau ),$ and total impurity susceptibility 
$\chi (\tau )$ for $U=2.3$~eV and $I=0.9$~eV are shown on the insets of Fig.~%
\ref{fig:chi_imp}. One can see that the dependence $\chi _{e_{g}}(\tau )$ on
imaginary time is more flat than $\chi _{t_{2g}}(\tau )$. This fact reflects
the formation of local moments for $e_{g}$ electrons, which would correspond
to fully time-independent $\chi _{e_{g}}(\tau )$. Switching off $I$
suppresses susceptibility at the flat parts, destroying therefore local
moments.

\begin{figure}[tbh]
\centering                           
\includegraphics[clip=true, width=0.33\textwidth,
angle=270]{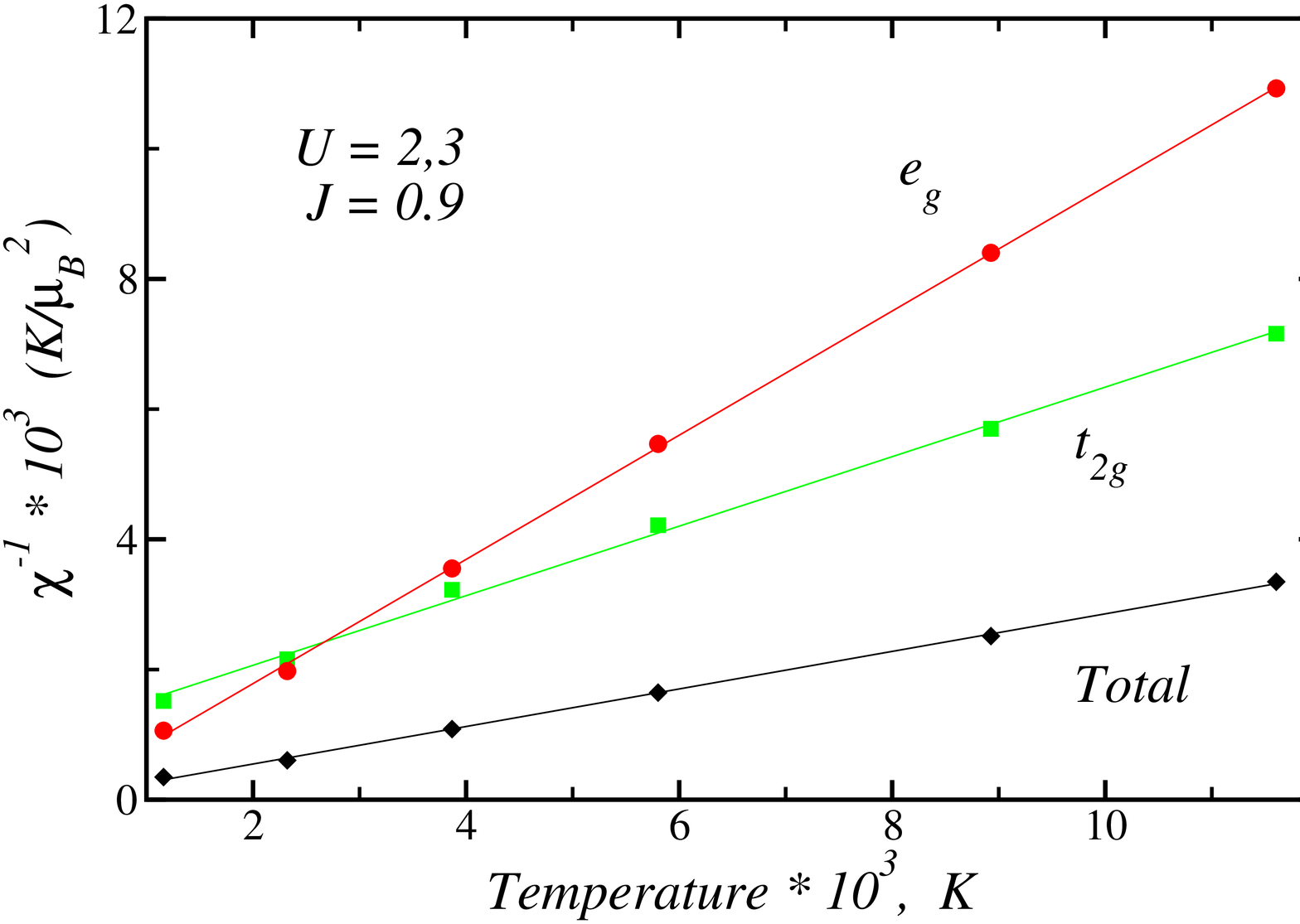}
\caption{(Color online) Temperature dependence of inverse of the local spin
susceptibility. Total spin susceptibility and orbital resolved contributions
are presented by (black) diamonds, (green) squares ($t_{2g}$) and (red)
circles ($e_{g}$), respectively. Lines are least-square fitting to the
original data.}
\label{fig:chi_vs_T}
\end{figure}

The observed behavior as a function of imaginary time is also reflected as a
function of real frequency (Fig.~\ref{fig:chi_imp}). One can see that flat
part of the imaginary-time dependence of the susceptibilities yields peak in
the real frequency dependence, which is mostly pronounced for $e_{g}$
states. The peak contributions are similar to the frequency dependence of
susceptibility of an isolated spin $p$ (note neglection of spatial
correlations in DMFT), $\chi (i\omega _{n})=g^{2}\mu
_{B}^{2}p(p+1)/(3T)\delta _{n,0}$ and show presence of local moment for $%
e_{g}$ states. For $t_{2g}$ states we observe mixed behavior with peak
contribution transferred from $e_{g}$ states via Hund exchange (see Sect.
IV) and incoherent background, originating from $t_{2g}$ itinerant states.
This peaky contribution to susceptibilities disappear with switching off $I$%
, which shows once more that Hund exchange is the major source of the local
moment formation.

One of the most transparent characteristic features of the local moment
formation is the fulfillment of the Curie-Weiss law for the temperature
dependence of the susceptibility. In particular, in the limit of local
moments the magnetic moment $\mu _{\text{CW}}$ extracted from Curie-Weiss
law is expected to be approximately equal to the magnetic moment in the
symmetry-broken phase. The obtained temperature dependence of the local
(impurity) susceptibilities is shown in Fig.~\ref{fig:chi_vs_T} (the
temperature dependence of lattice susceptibilities will be presented
elsewhere). One can see that the inverse susceptibility of $e_{g}$ states
obeys Curie law with $p_{\text{CW}}(e_{g})=0.52$. The inverse susceptibility
of $t_{2g}$ states also shows approximately linear temperature dependence
with $p_{\text{CW}}(t_{2g})\simeq 0.7$. The Curie law for the total
susceptibility yields $p_{\text{CW}}=1.16$ (the corresponding Curie constant 
$\mu _{eff}^{2}\equiv g^{2}\mu _{B}^{2}p_{\text{CW}}(p_{\text{CW}}+1)=10\mu
_{B}^{2}$) is in good agreement with experimental data and earlier
calculations of the lattice susceptibility in the paramagnetic phase~\cite%
{KatsPRL}. Note close proximity of obtained value $p_{\text{CW}}(e_{g})$ to
1/2.


\section{Effective model}

The formation of local moments by $e_{g}$ electrons makes the model~(\ref%
{Hdd}) reminiscent of the multi-band generalization of $s$-$d$ exchange
model, supplemented by Coulomb interaction in $t_{2g}$ bands. The $s$-$d$
model was first suggested by Shubin and Vonsovskii to describe magnetism of
rare-earth elements and some transition-metal compounds~\cite{Shubin}.
Differently to its original formulation, both itinerant and localized states
in the model~(\ref{Hdd}) correspond to $d$-electrons, with the $t_{2g}$ and $%
e_{g}$ orbital symmetry, respectively, and Coulomb interaction in $t_{2g}$
band is present.

Similarly to the diagram technique for the $s$-$d$ model~\cite{Izyumov}, the
contribution of the `wide band' $t_{2g}$ electrons can be treat
perturbatively. Moreover, we can integrate out electronic degrees of freedom
for $t_{2g}$ band and pass to purely bosonic model in a spirit of Moriya
theory. Specifically, we introduce new variables $\mathbf{t}_{q}^{m}$ for
spins of $t_{2g}$ electrons by decoupling interaction terms in $H_{t_{2g}}$
via Hubbard-Stratonovich transformation and summing contributions from
different spin directions (the double counted terms are supposed to be
subtructed). We treat only magnetic terms of the interaction, since we are
interested in magnetic properties. The Lagrangian, which is obtained by
Hubbard-Stratonovich transformation after expansion in the Coulomb
interaction between $t_{2g}$ states and Hund exchange can be represented in
the form 
\begin{widetext}
\begin{eqnarray}
L &=&L_{e_{g}}+\sum\limits_{q,mm^{\prime }}[R_{mm^{\prime }}^{-1}\mathbf{t}%
_{q}^{m}\mathbf{t}_{-q}^{m^{\prime }}-\Pi _{q}^{mm^{\prime }}(\mathbf{t}%
_{q}^{m}+2I\mathbf{S}_{q}\mathbf{)(t}_{-q}^{m^{\prime }}+2I\mathbf{S}_{-q})]
\notag \\
&&+\sum\limits_{q_{i},m_{i}}\Lambda _{q_{1}q_{2}q_{3},abcd}^{mm^{\prime
}m^{\prime \prime }m^{\prime \prime \prime }}(\mathbf{t}_{q_{1}}^{m}+2I%
\mathbf{S}_{q_{1}})_{a}(\mathbf{t}_{q_{2}}^{m^{\prime }}+2I\mathbf{S}%
_{q_{2}})_{b}(\mathbf{t}_{q_{3}}^{m^{\prime \prime }}+2I\mathbf{S}%
_{q_{3}})_{c}  \notag \\
&&\times (\mathbf{t}_{-q_{1}-q_{2}-q_{3}}^{m^{\prime \prime \prime }}+2I%
\mathbf{S}_{-q_{1}-q_{2}-q_{3}})_{d}  \label{HH}
\end{eqnarray}
where $R_{mm^{\prime }}=U\delta _{mm^{\prime }}+J(1-\delta _{mm^{\prime }}),$
the sums over band indices are taken over $t_{2g}$ states only, $%
a,b,c,d=x,y, $ or $z,$,
\begin{eqnarray}
\Pi _{q}^{mm^{\prime }} &=&-\sum\limits_{k}G_{k}^{mm^{\prime
}}G_{k+q}^{m^{\prime }m},  \notag \\
\Lambda _{q_{1}q_{2}q_{3},abcd}^{mm^{\prime }m^{\prime \prime }m^{\prime
\prime \prime }} &=&\frac{1}{16}\text{Tr}(\sigma ^{a} \sigma ^{b} \sigma ^{c} \sigma ^{d})\sum\limits_{k}G_{k}^{mm^{%
\prime }}G_{k+q_{1}}^{m^{\prime }m^{\prime \prime
}}G_{k+q_{1}+q_{2}}^{m^{\prime \prime }m^{\prime \prime \prime
}}G_{k+q_{1}+q_{2}+q_{3}}^{m^{\prime \prime \prime }m}
\end{eqnarray}
\end{widetext}$G_{k}^{mm^{\prime }}$ is the matrix of the (interacting) $%
t_{2g}$ electron Green functions, and we use the $4$-vector notations $%
q=(i\omega ,\mathbf{q})$ etc. Due to non-quasiparticle nature of $e_{g}$
electrons, the interaction acting on $e_{g}$ electrons and mixed $e_{g}$-$%
t_{2g}$ terms in the interaction need not be decoupled; the former supposed
to be accounted within a non-perturbative approach, e.g. DMFT, while the
latter are treated perturbatively. In dynamical mean-field theory quantities 
$\Pi _{q}^{mm^{\prime }}$ and $\Lambda _{q_{1}q_{2}q_{3},abcd}^{mm^{\prime
}m^{\prime \prime }m^{\prime \prime \prime }}$ for generic momenta are the
functions of frequencies only.

The Lagrangian (\ref{HH}) can be viewed as the generalization of the
standard $\phi ^{4}$ model of the magnetic transition of itinerant electrons 
\cite{Moriya,Hertz} to the case of presence of nearly local moments. For the
susceptibilities of $t_{2g}$ and $e_{g}$ electrons, and mixed $t_{2g}$-$%
e_{g} $ susceptibility\ $\upsilon _{q}$ we obtain up to second order in $I$ 
\begin{widetext}
\begin{eqnarray}
&&\left( 
\begin{array}{cc}
(R\chi _{q,t_{2g}}+I)R & \upsilon _{q}R \\ 
\upsilon _{q}R & \chi _{q,e_{g}}%
\end{array}%
\right) 
\begin{array}{c}
=%
\end{array}
\notag \\
&=&\left( 
\begin{array}{cc}
(\chi _{q,t_{2g}}^{0})^{-1}+\Lambda \ast \chi _{t_{2g}}^{0}+4I^{2}\Lambda
\ast \chi _{e_{g}}^{0} & -2I\Pi _{q} \\ 
-2I\Pi _{q} & (\chi _{q,e_{g}}^{0})^{-1}-4I^{2}\Pi _{q}+4I^{2}\Gamma
^{(4)}\ast \chi _{t_{2g}}^{0}%
\end{array}%
\right) ^{-1}  \label{sus}
\end{eqnarray}
\end{widetext}where $\chi _{q,t_{2g}}^{0}=(R_{mm^{\prime }}^{-1}-\Pi
_{q}^{mm^{\prime }})^{-1}$ is the RPA spin susceptibility of $t_{2g}$ band, $%
\chi _{q,e_{g}}^{0}=\langle \mathbf{S}_{q}\mathbf{S}_{-q}\mathbf{\rangle }%
_{e_{g}}/3$ is the bare susceptibility of $e_{g}$ band, evaluated with $%
L_{e_{g}}$, 
\begin{eqnarray}
\Gamma _{q_{1}q_{2}q_{3}}^{(4),abcd} &=&\langle
S_{q_{1}}^{a}S_{q_{2}}^{b}S_{q_{3}}^{c}S_{-q_{1}-q_{2}-q_{3}}^{d}\mathbf{%
\rangle }_{e_{g}}\mathbf{-}\chi _{q_{1},e_{g}}^{0}\chi
_{q_{3},e_{g}}^{0}\delta _{q_{1},-q_{2}}  \notag \\
&&-\chi _{q_{1},e_{g}}^{0}\chi _{q_{2},e_{g}}^{0}(\delta
_{q_{2},-q_{3}}+\delta _{q_{1},-q_{3}})
\end{eqnarray}%
is the 4-spin Green function, $\ast $ denote the convolution of momenta-,
frequency, and band indices. Again, within DMFT the quantities $\chi
_{e_{g},q}^{0}$ and $\Gamma _{q_{1}q_{2}q_{3}}^{(4),abcd}$ are only
frequency dependent.

The form of the susceptibilities~(\ref{sus}) allows in particular to
understand the mechanism of fulfillment the Curie law for local
susceptibilities of both, $t_{2g}$ and $e_{g}$ electrons and their frequency
dependence. While $e_{g}$ electrons form local moments, $\chi _{e_{g}}^{0}$
becomes almost static and shows inverse linear temperature dependence,
similar to that obtained in the Heisenberg model. The contribution $%
4I^{2}\Pi _{q}$ corresponds to RKKY interaction and expected to be weakly
temperature dependent. Presumably small contribution $4I^{2}\Gamma
^{(4)}\ast \chi _{t_{2g}}^{0}$ can also add some linear in temperature
dynamic contribution to the inverse susceptibility of $e_{g}$ electrons.
Note that within DMFT this contribution is accounted only in average with
respect to momenta, and does not allow to resolve peculiar physics, which
arises due to contribution of small momenta (forward scattering). The
convolutions $\Lambda \ast \chi _{t_{2g}}$ and $4I^{2}\Lambda \ast \chi
_{e_{g}}$ determine the contributions to the susceptibility of $t_{2g}$
electrons from interaction within $t_{2g}$ band and between $t_{2g}$ and $%
e_{g}$ bands, respectively, and become also linear functions of temperature
similarly to the Moriya theory (where they correspond to the so called $%
\lambda $-correction). These contributions are however incoherent due to
complicated frequency dependence of $\Lambda .$ Finally, the terms $I\Pi
_{q} $ mix these two (coherent and incoherent) contributions to the
susceptibilities due to interorbital Hund exchange and Coulomb interaction
in $t_{2g}$ band.

Therefore, the model~(\ref{HH}) allows to understand main features of
frequency- and temperature dependence of susceptibilities, observed in the
DMFT solution. The derivation of the bosonic model and susceptibilities~(\ref%
{sus}) can further serve as a basis for obtaining non-local corrections to
the results of dynamic mean-field theory, e.g. in a spirit of dynamic vertex
approximation~\cite{DGA,DGA1}.

\section{Conclusion}

We have discussed the origin of the formation of local moments in iron,
which is due to the localization of $e_{g}$ electrons. In particular, we
observe non-Fermi liquid behavior in $e_{g}$, but not $t_{2g}$ band. This
mechanism is very similar to the concept of orbital selective Mott
transition, which was earlier introduced in Ref.~\onlinecite{Anisimov} for Ca%
$_{2-x}$Sr$_{x}$RuO$_{4}$. Although a possibility of a separate Mott
transition in narrow bands (in the presence of hybridization with a wide
band) was questioned by Liebsch~\cite{Liebsch}, the recent high-precission
QMC studies of the two band model have confirmed this possibility~\cite%
{Dongen}. In our case, obtained non-Fermi liquid behavior of $e_{g}$
electrons yields peak in the frequency dependence of spin-spin correlation
function and linear temperature dependence of the magnetic susceptibility of 
$e_{g}$ electrons with $p_{\text{CW}}=0.52,$ both being characteristic
features of local moments, formed in $e_{g}$ band.

The formulated spin-fluctuation approach allows to describe thermodynamic
properties in the spin symmetric phase. To describe symmetry broken phase,
as well as proximity to the magnetic transition temperature, nonlocal (in
particular long-range) correlations beyond DMFT are expected to become
important. These correlations are also likely to reduce the DMFT transition
temperature closer to its experimental value. Although the systematic
treatment of the non-local long-range correlations in the
strongly-correlated systems is applied currently mainly to the one-band
models~\cite{DGA,DGA1,Cellular,Slezak}, it was shown recently that even for
the three-dimensional systems nonlocal corrections substantially reduce the
magnetic transition temperature from its DMFT value~\cite{DGA1}. Future
investigations of nonlocal corrections in multi-band models, together with
evaluation of thermodynamic properties, have to be performed.

The presented approach can be also helpful to analyse the electron structure
of $\gamma $-iron and mechanism of the structural $\alpha -\gamma $
transformation of iron~\cite{gammaFe}. Existing approaches to this problem
often start from the Heisenberg model, where the short-range magnetic order
in $\gamma $ phase was suggested as the origin of the $\alpha -\gamma $
transformation~\cite{trans}. This picture may need reinvestigation from the
itinerant point of view. The presented approach can be useful also for other
substances, containing both, local moments and itinerant electrons.

\section{Acknowledgement}

\bigskip The authors thank Jan Kune\v{s} for providing his DMFT(QMC)
computer code used in our calculations. Support by the Russian Foundation
for Basic Research under Grants No. RFFI-07-02-00041 and RFFI-07-02-01264-a,
Civil Research and Development Foundation together with the Russian Ministry
of Science and Education through program Y4-P-05-15, Federal Agency for
Science and Innovations under Grant No. 02.740.11.0217, the Russian
president grant for young scientists MK-1184.2007.2 and Dynasty Foundation,
the fund of the President of the Russian Federation for the support for
scientific schools NSH 1941.2008.2, the Program of Presidium of Russian
Academy of Science No. 7 \textquotedblleft Quantum microphysics of condensed
matter\textquotedblright , and grant 62-08-01 (by \textquotedblleft
MMK\textquotedblright , \textquotedblleft Ausferr\textquotedblright , and
\textquotedblleft Intels\textquotedblright ) is gratefully acknowledged.

\end{document}